\def\tsecompldate{8th April, 1995}
\def\tseprepno{Imperial/TP/94-95/26} %{Preliminary version}
\def\tseepreno{94-5\_26.tex} % {94-5\_xx.tex}
\def\tsehepphno{{\tt hep-ph/9504266}}

\def\tsetrue{T} \def\tsefalse{F} % see TeX book pp 206 and environs

\let\tsepaper=\tsefalse   % select paper mode, preprint is default
\let\tsenoteon=\tsefalse   % tse note system on
\let\tselse=\tsetrue      % label as (sec.eqn) otherwise (eqn)
\let\tseletter=\tsetrue  % select Letter style paper, A4 is default
\let\tsedevon=\tsefalse    % select development options on or off
\let\tseepsfon\tsefalse   % select reading in of eps figures
\if\tsefalse\tseepsfon\documentstyle[12pt]{article}
\else\documentstyle[12pt,epsf]{article}
\fi
\if\tsetrue\tsedevon \newcommand{\tsedevelop}[1]{{#1}}
\else \newcommand{\tsedevelop}[1]{{}}
\fi

\if\tsetrue\tsepaper 
\else 
\fi

\tsedevelop{\typeout{*** T.S.E. Development mode on ***}}

\if\tsetrue\tseepsfon 
\else 
\fi

\if\tsetrue\tsepaper 
\else 
\fi

\if\tsetrue\tsepaper  
\fi

\if\tsetrue\tseletter{
\typeout{          as for Letter paper ---}
}\else
\fi
\if\tsetrue\tselse{\makeatletter

\@addtoreset{equation}{section}
\makeatother
\fi
\if\tsetrue\tselse
\typeout{--- Equations labeled as (section.equation) ---}
\fi

\newcommand{\auth}[1]{#1}
\newcommand{\journal}[1]{#1}
\newcommand{\vol}[1]{{\bf #1}}
\newcommand{\yr}[1]{(#1)}

\newcommand{\ttitle}[1]{{\it #1}}
\newcommand{\pretitle}[1]{{\it #1}}
\newcommand{\inproctitle}[1]{``#1''}

\if \tsedevon\tsefalse \newcommand{\tselea}[1]{\label{#1}}
\else \newcommand{\tselea}[1]{\label{#1} \\
& \mbox{ } & \mbox{ (#1) } \nonumber }
\typeout{--- Renewing tselabel command ---}
\fi
\newcommand{\tseleq}[1]{\tsedevelop{\mbox{  (#1) }}\label{#1}}

\newcommand{\tbib}[1]{\bibitem{#1}\tsedevelop{ [#1] }}
\newcommand{\tref}[1]{(\ref{#1}\tsedevelop{-#1})}

\newcommand{\tcite}[1]{\cite{#1}\tsedevelop{ [#1] }}

\newcommand{\tnote}[1]{\if\tsetrue\tsenoteon \footnote{#1} \fi}
\if\tsetrue\tsenoteon{
\typeout{--- Tim Footnotes Included ---}
}\else \typeout{--- Tim Footnotes Excluded ---}
\fi

\if\tsetrue\tseepsfon \newcommand{\tseepsffile}[1]{\epsffile{#1}}
\newcommand{\tseepsfxsize}[1]{\epsfxsize=#1}
\else \newcommand{\tseepsffile}[1]{{}}
\newcommand{\tseepsfxsize}[1]{{}}
\fi

\newcommand{\tcaption}[2]{
\if\tsetrue\tsepaper \vspace{5cm} \caption{  }
\else
\vspace{#1} \caption{#2}
\fi }

\newcommand{\half}{\frac{1}{2}}
\newcommand{\bea}{\begin{eqnarray}}
\newcommand{\eea}{\end{eqnarray}}
\newcommand{\beq}{\begin{equation}}
\newcommand{\eeq}{\end{equation}}
\newcommand{\nnel}{\nonumber \\ {}}
\newcommand{\ennlpp}{ {} } % {\nonumber \\ && {}} set blank if not preprint
\typeout{--- Equation break set for wide text ---}
\newcommand{\npagepub} {\if\tsetrue\tsepaper{\newpage }\fi}
\if\tsetrue\tsepaper{
\typeout{--- Page Breaks for Pub. Version ON ---} }\else{
\typeout{--- Page Breaks for Pub. Version OFF ---} }\fi

\newcommand{\quarter}{\frac{1}{4}}
\newcommand{\ra}{\rightarrow}

\newcommand{\dkj}[1]{\int dk_1 \ldots dk_{j}}

\def\gsim{\raise.3ex\hbox{$>$\kern-.75em\lower1ex\hbox{$\sim$}}}
\def\lsim{\raise.3ex\hbox{$<$\kern-.75em\lower1ex\hbox{$\sim$}}}
\def\gsim{\raise.3ex\hbox{$>$\kern-.75em\lower1ex\hbox{$\sim$}}}
\def\lsim{\raise.3ex\hbox{$<$\kern-.75em\lower1ex\hbox{$\sim$}}}

\newcommand{\dash}[1]{{#1}^\prime{}}

\newcommand{\sigmabar}{\bar{\sigma}}
\newcommand{\epsilonbar}{\bar{\epsilon}}
\newcommand{\psibar}{\bar{\psi}}

\newcommand{\set}[1]{\mbox{$\{ #1 \}$}}

\newcommand{\permb}{\sum_{perm. \{b\} }}
\newcommand{\permbnc}{\sum_{perm. \{b\} | NC}}
\newcommand{\Real}{{\Re e}}

\newcommand{\twf}{{\cal W}}
\newcommand{\twft}{{\tilde{\twf}}}
\newcommand{\PP}{{\rm PP}}
\begin{document}

\typeout{--- Title page start ---}

\if\tsefalse\tsepaper \thispagestyle{empty}\fi

\renewcommand{\thefootnote}{\fnsymbol{footnote}}

\begin{flushright}
\tseprepno \\
\tsehepphno \\
\tsecompldate \\
\tsedevelop{ (\LaTeX -ed on \today ) \\}
%{next preprint number}\\
\end{flushright}
\vskip 1cm

\begin{center}
{\Large\bf Thermal Green Functions at Zero
Energy}\footnote{Available in {\LaTeX} format
through anonymous ftp from
{\tt ftp://euclid.tp.ph.ic.ac.uk/papers/\tseepreno},
or on WWW in {\LaTeX}  and postscript formats at
{\tt http://euclid.tp.ph.ic.ac.uk/Papers/index.html} }
\vskip 1.2cm
{\large\bf T.S. Evans\footnote{E-mail: {\tt T.Evans@IC.AC.UK};
WWW: {\tt http://euclid.tp.ph.ic.ac.uk/$\sim$time}}}\\
Blackett Laboratory, Imperial College, Prince Consort Road,\\
London SW7 2BZ  U.K.
\end{center}

%\if\tsetrue\tsepaper{ \begin{center}
%  Tel: U.K.-171-594-7837 \\  Fax: U.K.-711-594-7777 \\
%  \mbox{  }\\   PACS: 11.10-z
%  \end{center} }\fi

\npagepub

\vskip 1cm
\begin{center}
{\large\bf Abstract}
\end{center}
%text of abstract here

The thermal expectation values of all possible bosonic generalised
retarded functions evaluated at zero energy are studied.   The
relationship of such functions  to calculational schemes, technical
problems and physical applications is outlined.  It is then shown
that all generalised retarded functions constructed from any one set
of bosonic fields are equal  at zero energy.  This is done
completely generally and is not limited to any approximation scheme
such as perturbation theory.

\vskip 1cm

%See pp.175
\renewcommand{\thefootnote}{\arabic{footnote}}
\setcounter{footnote}{0}

\npagepub

\typeout{--- Main Text Start ---}

\section{Introduction}

There are many types of Green function which appear in thermal field
theory.  The retarded and advanced Green functions
\tcite{RAinRTF,Ge}  are merely a subset of the much more numerous
GRF (generalised retarded functions)
\tcite{Arakietc,TSEnpt,TSEllwi}.  The GRF as a whole are most easily
calculated in ITF (the Imaginary-Time Formalism or Matsubara method)
\tcite{Ge,Raybook,LvW,ITF},  and they are the only ones calculated
directly in ITF \tcite{TSEnpt,TSEllwi}.   Conversely calculations in
ITF invariably produce GRF.   If a real-time formalism is used
\tcite{RAinRTF,Raybook,LvW} of all the GRF it is only known how to
extract the retarded and advanced functions
\tcite{RAinRTF,TSEnpt,TSEllwi}.   However the precise method used to
calculate the GRF is of no relevance to the manipulations performed
in later sections.  Calculational schemes will only be discussed
with respect to the motivation and background to the problem of
bosonic GRF at zero energy, to which we now turn.

The zero energy Green functions have proved to have a much more
complicated behaviour than their zero temperature
counterparts.  This has been seen when specific examples are
calculated \tcite{ZEdiag}.   This is due
to the many cuts running across the zero energy point \tcite{We}.
By looking at simple perturbative examples \tcite{We}, one can
quickly see that for a thermal Green function there are {\em always}
cuts running across the zero energy point (though not always at the
lowest order of perturbation theory).  Physically this is due to Landau
damping processes.

The zero energy Green functions are of great physical importance.
At zero temperature they are generated by the lowest term in a time
derivative expansion of the effective action.  However, how can one
make derivative expansions of the effective action in thermal field
theories when zero-energy thermal Green functions show a marked
dependence \tcite{ZEdiag} on the order in which spatial and
time derivative expansions are made?   Conversely, precisely what
sort of thermal Green functions (retarded, advanced, GRF, time
ordered, thermal Wightman, etc.) are generated by the effective
action derivative expansions used in the literature?

A related issue is that the zero energy Green functions ought by
default to be pure real if one is thinking in terms of effective
action expansions around a simple stable vacuum configuration.
However, from a mathematical point of view, why
should these thermal Green functions be real when they are generally
complex near zero energy due to cuts which are present in the thermal case?

Another question involves direct calculations of Green functions at
zero energy in ITF.  In this case all the external energies are set
directly equal to the zero discrete Euclidean energy value.
 Why though do we pay no heed to the large number of cuts which are
known to pass across the zero energy point?  In turn, how does this
static ITF calculation relate to the zero energy limits of the
different GRF \tcite{ZEdiag}.  Looking at it from another point of
view, why, when using contour methods to evaluate ITF energy sums,
does it not matter which side of the cuts the discrete energy point
at zero energy is put?

In the following study of the zero energy limit of the GRF at finite
temperature, we will use the approach used in \tcite{TSEnpt,TSEllwi}
to study general thermal Green  functions.  This uses the
definitions of the Green functions as expectation values of fields
and the fundamental relations between them such as the KMS
condition.  These relations only involve factors of $e^{-\beta p}$
where $p$ is an external energy.  Thus looking at the low energy
limit is equivalent to looking at the infinite temperature limit and the
results are valid for either viewpoint.  No other scales are
relevant to the results of this analysis!  A big advantage of such
analysis is that the results are completely general, one can examine
any type of field, and they apply to both the full Green functions and
to any sensible  approximation to the Green functions.   All that is
required is that the fundamental relations between Green functions
are satisfied in what ever approximation scheme is used.  As the KMS
condition is  equivalent to the definition of what thermal
temperature field theory is, there is no problem with any useful
approximation.

In section two, we recall some of the basic results of \tcite{TSEnpt}
to be used in the subsequent proofs.  This will also serve to
establish the notation used.  The most general proof given here
involves a hideous shuffling and partitioning of subscripts and
subsubscripts.  For this reason in the following sections, simpler
cases, which illustrate the basic principles involved, are considered
in order of increasing complexity, before the final most general and
undignified proof is presented.

\section{Thermal Generalised Retarded Functions}

Throughout this paper all quantities are being measured from  the
rest-frame of the heat bath.  Any dependence on spatial coordinates
or three-momentum is not written explicitly as this does not effect
the arguments given here.  This is because the thermal boundary
conditions only involve the external energies.   Variables such as
$p,k$ etc. refer to real Minkowskii energies and not energy-momentum
four-vectors.  When we have continued to the complex energy plane,
we will denote general complex energies by $z$.

We will consider a set of $N$ bosonic fields labeled $\phi_1, \ldots,
\phi_N$ as in principle they may be distinct types of field
(e.g.\ different components of scalar or gauge fields) as well as carrying
their own time arguments.  These appear in
many different orders so it is convenient to use subscripts
$\phi_{b_j}$ where the $N$ subscripts $b_j$ ($j=1\ldots N$) are some
permutation of the integers from $1$ to $N$.  We will then need to
shuffle these subscripts round further to enable different terms to
be compared.  This means that the subsubscripts are now manipulated.
 Any reference to subsubscripts outside the range $j=1\ldots N$ are
to be understood to be modulo $N$ i.e.\ $b_0=b_N$ etc.

In ITF one calculates expectation values of Euclidean time ordered
fields whose time arguments lie between $0$ and $- i \beta$.
The result of a calculation of an N-point function in ITF is given
by some function, say $\Phi$, evaluated at discrete Euclidean energy
points, i.e.\ $\Phi ( \{  z_j=2\pi \nu_j / \beta \} )$ ($\nu_j$ are integers
for bosonic fields).   To look
at the region around the zero energy point we need to make an analytic
continuation.  Analytic
continuation to general complex energies, $\{ z \}$, can be
performed uniquely if one chooses
certain behaviour at large $|z|$ \tcite{TSEnpt,BM}.  In many
practical perturbative calculations this is usually trivial.
The resulting
function, $\Phi(\{z \})$, is found to represent the analytic
continuation from real to complex energies of what
are called GRF (Generalised
Retarded Functions) \tcite{Arakietc,TSEllwi}.  The form found is
\tcite{TSEnpt,TSEllwi}
\begin{eqnarray}
\Phi(\{z\} ) &=&
\left(\frac{-1}{2\pi}\right)^{N-1} \dkj{N}  \; \delta( \sum_{j=1}^N k_j )
. \nnel &&
\permb \twf(b_{1},b_{2},...,b_{N} ; \{k\}) .
\prod_{j=2}^{N} \frac{i}{B^{N}_j}
\tselea{ePhidef}
\end{eqnarray}
where the first sum takes $\{b_j\}$ through all
permutations of the numbers $(1,2,...,N)$ and
\begin{eqnarray}
B_i^j &=& \sum_{l=0}^{|j-i|_N} (z_{b_{l+i}} - k_{b_{l+i}} ),
\tselea{eBdef1}
\end{eqnarray}
To write $\Phi$ in this way, an $N$-th
redundant complex energy variable has been introduced
defined through the constraint
\begin{equation}
\sum_{j=1}^N z_j = 0.
\tseleq{econstraint}
\end{equation}

The $\twf(b_{1},b_{2},...,b_{N} ; \{k\}) $ are the thermal Wightman
functions in energy space defined for pure bosonic fields to be
\begin{eqnarray}
\lefteqn{\twft (b_{1},b_{2},...,b_{N} ; \tau_1,\tau_2,...,\tau_N) =} \nnel &&
Tr\{e^{-\beta H}
\phi_{b_1}(\tau_{b_1})\phi_{b_2}(\tau_{b_2})...\phi_{b_N}(\tau_{b_N}) \}
/ Tr\{e^{- \beta H}\}. \ennlpp
\tselea{entwf} \\
\lefteqn{ \twft (b_{1},b_{2},...,b_{N} ; \tau_1,\tau_2,...,\tau_N)=}
\nnel &&
(2\pi)^{-N} \left( \prod_{j=1}^N \int_{-\infty}^{\infty} dp_j
e^{-i p_j \tau_j} \right) .
%\nnel && \mbox{     }
\twf(b_{1},b_{2},...,b_{N} ; p_1,p_2,...,p_N)
\tselea{eq:N8}
\end{eqnarray}
The $\{\tau\}$ are complex times.
{}From this definition in terms of the trace and Boltzman factor
we find a fundamental property of these Green functions, a
generalisation to $N$-point functions of the well known KMS
(Kubo-Martin-Schwinger)
condition for two-point functions \tcite{Raybook,LvW,BM,ITF},
namely for pure bosonic functions
\begin{equation}
\exp(- \beta p_{b_1})
\twf (b_{1},b_{2},...,b_{N} ;p_1,p_2,...,p_N)=
\twf (b_{1},b_{2},...,b_{N} ;p_1,p_2,...,p_N) .
\tseleq{etcyclee}
\end{equation}

Immediately we can see that there are poles in the integrand of
\tref{ePhidef} for real external  energies.  This leads
to discontinuities in $\Phi(\set{z})$ at real energies (as $\twf \neq
0$ in general for the thermal case)
so we must specify how we approach the real energy
axes.  In fact we find that there are many different results
depending on how the axes are approached (many more than $N!$ in general
\tcite{TSEnpt,TSEllwi}).  We will specify which sides of the
cuts along the real energy axes are being considered by
setting the complex energies to
\beq
z_i=p_1+\epsilon_i, \; \; \; p_i,\epsilon_i \in \Re e, \; \;
|\epsilon_i| \ll 1
\tseleq{eepdef}
\eeq
To ensure that one is not sitting on
any cuts, and from \tref{econstraint}, the epsilons must satisfy:-

1. All sums of subsets of the epsilons must be non zero
\beq
\sum_{j=1}^{N} c_j \epsilon_j \neq 0 \; \;
\forall \; \{c_j\}=\{0,1 |0<\sum_{j=1}^{N}  c_j <N \}
\tseleq{eepcond1}
\eeq

2. The sum of all epsilons must equal zero
\beq
\sum_{j=1}^{N} \epsilon_j = 0
\tseleq{eepcond2}
\tref{eepcond2}
\eeq

It is the sign of the epsilons and the sign of all possible sums of
the epsilons which completely specify which side of the real energy cuts
of $\Phi$ function is being
studied.  Complex energy space is split into a large number of
regions, each bounded by real energy cuts and each region
corresponds to a unique GRF \tcite{TSEnpt,TSEllwi}.
We therefore define two sets of epsilons to be equivalent if they
have selected the same region of complex energy space and so are giving
precisely the same GRF.
\bea
\set{\epsilon} \equiv \set{\dash{\epsilon}} & \mbox{ iff } &
{\rm SGN}( \sum_{j=1}^{j=N} c_j \epsilon_j) =
{\rm SGN}( \sum_{j=1}^{j=N} c_j \dash{\epsilon}{}_j ) \nnel
&& \forall \; \{c_j\}=\{0,1 \; | \; 0<\sum c_j <N\}
\tselea{eepequiv}
\eea
where
\beq
{\rm SGN}(x) =
\protect\left\{
\begin{array}{cl}
+1, & x>0  \\
-1, & x<0
\end{array}
\right.
\tseleq{eSGN}
\eeq

An important subset of the generalised retarded functions comes from
just $2N$ different ways of choosing
these epsilons.
Suppose in \tref{eepdef} we pick out one epsilon,
and set it positive, and then set all the other epsilons
negative, say
$\epsilon_a=(N-1)\epsilon$, $\epsilon_{other}=-\epsilon$ where
$\epsilon$ is an infinitesimal positive quantity.
In this case, the ITF result after analytic continuation,
is found to be, in terms of real times,
\begin{eqnarray}
\Phi^{(N)}(\{t\};\epsilon_a>0, \epsilon_{other}<0)
&=&  R_a (\{t\})  ,
\tselea{enitfrr}
\end{eqnarray}
where $R_a$ is one of the $N$ retarded N-point functions.  For pure
bosonic fields they can be written as
\begin{eqnarray}
\lefteqn{R_a(t_1,t_2,...,t_N) = } \nnel &
\sum_{{\rm perm}\{a\}|a_N=a} &
\theta(t_{a_{N}}-t_{a_{N-1}}) \theta(t_{a_{N-1}}-t_{a_{N-2}})
\ldots \theta(t_{a_{2}}-t_{a_{1}}) .
\nnel
&& \; \; \; \;
[[\ldots [[\phi_a,\phi_{a_{N-1}}],\phi_{a_{N-2}}],\ldots ],
\phi_{a_{1}}] \ennlpp
\tselea{enptbr}
\end{eqnarray}
where $\phi_a=\phi_a(t_a)$ and the $\phi_a$ can be different bosonic fields.
The sum takes the $\{a_j\}$, $j=1$ to $N-1$,
through all permutations of the numbers $1$ to $N$ less the number
$a$.  The number $a$, which is the subscript on the $R$,
indicates that the $a$'th field has the largest time and we set $a_N=a$.

The $N$ advanced functions, $A_a$,
are obtained in an identical manner except
that the theta functions are reversed in time and an overall factor of
$(-1)^{N-1}$ is added.  This corresponds to
switching the signs of all the epsilon in \tref{ePhidef} terms so we have
\begin{eqnarray}
\Phi^N(\{t\};\epsilon_a<0, \epsilon_{other}>0)
&=& A_a (\{t\}).
\tselea{enitfar}
\end{eqnarray}

The retarded and advanced functions are merely a subset of the
generalised retarded functions \tcite{Arakietc,TSEnpt,TSEllwi}.
In this case the
various different analytic continuations of $\Phi$ to the real
energy axes form a definition of the generalised retarded functions
in energy space.  This in turn then gives a definition of the GRF in
terms of fields and theta functions in real time.

\section{Two-point functions}

The method to be used here to study zero energy N-point
Green functions is easily
illustrated for two-point functions.  Some well known
results are reproduced in a slightly more cumbersome notation
\tcite{LvW,BM,ITF}.  For $N=2$ we have
\bea
\Phi^{(2)}(z_1) &=&
\frac{1}{2\pi} \int dk_1 dk_2 \;  \delta(k_1+k_2)
\left[ \twf(12;k) \frac{i}{z_1-k_1 }   +
\twf(21;k) \frac{i}{z_2-k_2 } \right]
\tselea{e2ptPhi}
\eea
\beq
R^{(2)}(p_1) = \Phi(p_1 + i\epsilon), \; \; \; \; A^{(2)}(p_1) =
\Phi(p_1-i\epsilon)
\tseleq{e2ptRA}
\eeq
where $\epsilon$ without a subscript is an infinitesimal positive
real number, $p_1+p_2=0$, and $\epsilon_1+\epsilon_2=0$.
We use the usual representation
\beq
\frac{i}{p_j-k_j+i \epsilon_j} =
\frac{i \PP}{p_j-k_j} +
\theta_j \pi \delta (p_j-k_j)
\tseleq{ePPdelta}
\eeq
where $\theta_j= +1$ ($-1$) when $\epsilon_j>0$ ($<0$).  The $\PP$
indicates that the principal part is to be taken.

We now use \tref{ePPdelta} to split \tref{e2ptPhi} into two pieces
\bea
\Phi^{(2)}(z_1)=\Phi^{(2)}_0(z_1)+\Phi^{(2)}_1(z_1)
\eea
{}From this we find for the two-point functions
\bea
\Phi^{(2)}_0(p_1+ i \epsilon_1) =
\frac{1}{2\pi} \int dk_1 dk_2 \;  \delta(k_1+k_2)
& \left[ \right. &
\twf(12;k) \frac{i \PP}{p_1-k_1 }   +  \nnel && \left.
\twf(21;k) \frac{i\PP}{p_2-k_2 } \right]
\nnel
\Phi^{(2)}_1(p_1+i\epsilon_1) =
\half \int dk_1 dk_2 \;  \delta(k_1+k_2)
& \left[ \right. &
\twf(12;k) \theta_1\delta(p_1-k_1 ) +
\nnel && \left.
\twf(21;k) \theta_2\delta(p_2-k_2) \right]
\eea
The KMS condition \tref{etcyclee} tells us that
$e^{-\beta p} \twf (12;p) =
\twf (21;p)$.  If we now use this in the  zero energy limit, and use
the various relations between the energy variables such as
\beq
\theta_1+\theta_2=0
\tseleq{e2ptPsi}
\eeq
we find
\bea
\Phi^{(2)}_0(p_1+i\epsilon_1)&=&
\frac{1}{2\pi} \int dk
(\twf(12;k) -\twf(21;k)) \frac{i\PP}{-k }
\nnel &=&
R^{(2)}(p=0) = A^{(2)}(p=0)
\\
\Phi^{(2)}_1(p_1+i\epsilon_1) &=& 0
\eea
This is a well known result \tcite{BM,LvW,ITF} that the imaginary part of
two-point bosonic functions is zero at zero energy and
arbitrary three-momenta.  Physically this result is important for the
study of electric and magnetic static field screening in a gauge theory.

\section{Three-point functions}

For three-point functions there are three retarded, $R_a$,
and three advanced, $A_a$, functions.  They illustrate why the general
$N$-point case is not nearly as simple as the two-point example.
{}From \tref{ePhidef}, the three-point GRF is given by
\bea
\lefteqn{\Phi^{(3)}(z_1,z_2,z_3=-z_1-z_2) =
\frac{1}{4 \pi^2} \int dk_1 dk_2 dk_3 \;
\delta(k_1+k_2+k_3) . }\nnel
&& \sum_{{\rm perm.}123}
\twf (123;k_1,k_2,k_3) \frac{i }{(z_2+z_3-k_2-k_3)} \frac{i }{(z_3-k_3)}
\tselea{e3ptGRF}
\eea
where the sum is over all permutations of the indices.
The retarded and advanced
products are obtained by considering the 6 regions of complex energy
space separated by the cuts along the real energy axes.
We set $z_a = p_a+i\epsilon_a$
where $p_a, \epsilon_a \in \Real $, $\epsilon_a$ is a
non-zero infinitesimal and $\theta_a=+1$ $(-1)$ if $\epsilon_a>0$
$(<0)$.  We then split $\Phi^{(3)}$ into three pieces using
\tref{ePPdelta} and find
\beq
\Phi^{(3)} =\Phi^{(3)}_0 + \Phi^{(3)}_1 + \Phi^{(3)}_2
\eeq
where
\bea
\lefteqn{\Phi^{(3)}_0(p_1+i\epsilon_1,p_2+i\epsilon_2) =
\frac{1}{4 \pi^2} \int dk_1 dk_2 dk_3 \;
\delta(k_1+k_2+k_3) . } \nnel
&& \sum_{{\rm perm.} 123}
\twf (123;k_1,k_2,k_3) \frac{i \PP}{(p_2+p_3-k_2-k_3)}
\frac{i \PP}{(p_3-k_3)}
\tselea{e3pt0d}
\eea
\bea
\lefteqn{\Phi^{(3)}_1(p_1+i\epsilon_1,p_2+i\epsilon_2) =
\frac{1}{4 \pi}  \int dk_1 dk_2 dk_3 \;
\delta(k_1+k_2+k_3) .} \nnel
& \sum_{{\rm perm.}123} &
\twf (123;k_1,k_2,k_3)\left[ \frac{i \PP}{p_2+p_3-k_2-k_3}
\theta_3 \delta(p_3-k_3) + \right.
\nnel
&&
\left. \theta_{23} \delta(p_2+p_3-k_2-k_3)
\frac{i \PP }{p_3-k_3} \right]
\tselea{e3pt1d}
\eea
\bea
\Phi^{(3)}_2(p_1+i\epsilon_1,p_2+i\epsilon_2)
&=& \quarter \sum_{{\rm perm.}123}
\twf (123;p_1,p_2,p_3) \theta_{23} \theta_3
\tselea{e3pt2d}
\eea
where $\theta_{23}=+1$ ($-1$) if $\epsilon_2+\epsilon_3>0$ ($<0$) etc.
Clearly $\Phi^{(3)}_0$ is independent of
the which argument we choose to be retarded or advanced, while the
remaining two are not obviously independent of which retarded or
advanced function we are looking at.

Looking at the $\Phi^{(3)}_2$ term
we can show that it is independent of the index $a$ at zero external
energy by using the KMS
condition for bosonic functions \tref{etcyclee}.
For three-point functions at zero energy this is
\beq
\twf (312;0)=\twf (123;0)=\twf (231;0), \; \;
\twf (321;0)=\twf (213;0)=\twf (132;0) .
\tseleq{e3ptkms}
\eeq
We also have that
\beq
\theta_{23} \theta_3 + \theta_{31} \theta_1 + \theta_{12} \theta_2 = -1
\tseleq{e3ptPsi2}
\eeq
whatever $a$ is chosen.  This then gives
\beq
\half \left(R^{(3)}_a(0,0) + A^{(3)}_a(0,0) \right) =
\Phi^{(3)}_0 - \quarter \left( \twf (123;0)+\twf (321;0) \right)
\mbox{   independent of $a$.}
\eeq

The terms with neither all delta functions nor all principal parts are
the difficult ones in the general case, and the $\Phi^{(3)}_1$ term
illustrates how such terms are manipulated.  The principal part
integrands have to be integrated over and it is difficult to say what
these give
in general.  It also means we can not put all the integration
variables to zero in any one term, as we could with $\Phi^{(3)}_2$
and so the equality \tref{e3ptkms} between bosonic
$N$-point functions at zero energy can not be immediately used.
One proceeds by
collecting terms with the same delta functions. Consider
terms containing $\delta(p_3-k_3)$.  This comes from two places,
the term shown explicitly in \tref{e3pt1d} and one from the second
delta function term where we have
cycled the indices $123 \rightarrow 231$.  Thus we can rewrite the
second term in the square bracket in \tref{e3pt1d} as
\bea
\lefteqn{\Phi^{(3)}_1(p_1+i\epsilon_1,p_2+i\epsilon_2) = \frac{1}{4\pi}
\int dk_1 dk_2 dk_3 \;
\delta(k_1+k_2+k_3)} . \nnel
& \sum_{{\rm perm.} 123} &
\left[ \theta_3 \delta(p_3-k_3)
\twf (123;k_1,k_2,k_3) \frac{i \PP}{p_2+p_3-k_2-k_3}  + \right.
\nnel && \left.
\theta_{12} \delta(p_1+p_2-k_1-k_2)
\twf (312;k_1,k_2,k_3) \frac{i \PP }{p_2-k_2}
\right]
\tselea{e3pt1d2}
\eea
Now at zero external energy, the delta function does at least set
one of the integration variables to zero.  Together with the KMS condition
\tref{etcyclee} for pure bosonic
functions in this case, we find
\bea
\twf (312;k_1,-k_1,0) &=& \twf (123;k_1,-k_1,0)
\\
\frac{i \PP}{p_2-k_2} &=& - \frac{i \PP }{p_2+p_3-k_2-k_3}
\\
\theta_3+\theta_{12} &=& 0
\tselea{e3ptPsi1}
\eea
and hence we see that
\beq
\half \left(R^{(3)}_a (p=0) - {A^{(3)}_a} (p=0) \right)
= \Phi^{(3)}_1 = 0  .
\eeq
This leaves us with
\beq
R^{(3)}_a(0,0) = A^{(3)}_a(0,0)  =
\Phi^{(3)}_0 - \quarter \left( \twf (123;0)+\twf (321;0) \right)
\; \; \mbox{   independent of $a$.}
\eeq
So all the three-point GRF made out of the same set of bosonic fields
are equal at zero energy.

\section{N-point functions}

Using \tref{ePPdelta} split up $\Phi$,
\beq
\Phi(\{p+i \epsilon\} ) = \sum_{l=0}^{N-1} \Phi_l(\{p+i \epsilon\} )
\eeq
where $\Phi_l$ has $l$ of the $B$ terms of $\phi$ in \tref{ePhidef}
replaced by delta functions and the other $N-l-1$ factors of $B$ are
replaced by principal parts.  Thus
\begin{eqnarray}
\lefteqn{\Phi_l(\{z_a=p_a + i \epsilon_a \} ) } \nnel
&=&
\left(\frac{-1}{2}\right)^{N-1}
\left(\frac{1}{\pi}\right)^{N-l-1}
\dkj{N}  \; \delta( k_1+k_2+\ldots+k_N)
. \nnel &&
\permb \sum_{ \{x\} \in X_0  }
\twf (b_{1}b_{2}...b_{N};\{k\}) .
\frac{i \PP}{B^{N}_2} \ldots
\frac{i \PP}{B^{N}_{x_1-1}}
\psi_{x_1}^N \delta(B^{N}_{x_1} )
\frac{i \PP}{B^{N}_{x_1+1}}
%\overline{ \frac{i \PP}{B^{N}_{x_1}} }
\ldots
\nnel
&& \; \; \; \; \; \; \; \; \mbox{       }
\ldots \psi_{x_l}^N \delta(B^{N}_{x_l} )
\ldots \frac{i \PP}{B^{N}_N}
\tselea{enptitfl}
\end{eqnarray}
where we are replacing the $x_i$-th $B$ factor, $B^N_{x_i}$, by a
delta function and the other $B$ factors are replaced by their
principal parts.  The effective definition of $B$ is now
\begin{eqnarray}
B_i^j = B(i,j; \{ p-k \}) &=&
\sum_{l=0}^{|j-i|_N} (p_{b_{l+i}} - k_{b_{l+i}} ),
\tselea{eBdef2}
\end{eqnarray}
as the infinitesimal
imaginary parts of the external energies are dealt with
explicitly.  This replacement of
the $B$ terms by  principal parts and delta functions
is done in all possible ways, and this performed by the
$\sum_{ \{x\}}$. This sum takes the $x_j$ variables through all
values between 1 and $N$ keeping the $x$'s in ascending order i.e.\
$\{ x \} \in X_0$ where
\beq
X_j= \{ \{ x \} | 1+j=x_0 < x_1< \ldots <x_l \leq
N+j \equiv x_0 -1 \}.
\tseleq{eXdef}
\eeq
It is implicit through out this work that whenever needed, indices,
such as the $x$'s and $b$'s, are periodic so that $x_j \equiv x_j + N$.
The signs of the small epsilon regulating terms are encoded by the
$\psi=\pm 1$ factors which are defined to be
\begin{eqnarray}
\psi_j^m&=&
\protect\left\{
\begin{array}{l}
+1  \\
-1
\end{array} \right\}
\mbox{ if } \sum_{l=0}^{m-j} \epsilon_{b_{j+l}}
\protect\left\{
\begin{array}{l}
 >0  \\
 <0
\end{array} \right.
\end{eqnarray}

The idea is to exploit the delta functions in \tref{enptitfl}.   The
indices are cycled as a subset of the sum over all permutations of
indices.  This means that terms with similar delta functions and
principal parts appear $l+1$ times.  In fact one can quickly see that by
using the delta functions,  the KMS condition and the fact that we will
put $\{ p \}$ to zero, the only complicated part comes from the $\psi$
terms.

It is more useful, in view of the cyclic properties of the thermal
Wightman functions, to write out this part of the permutation of $b$
mindices explicitly.  It is possible to write $\Phi_l$ in a slightly
different form, namely
\begin{eqnarray}
\lefteqn{\Phi_l(\{z_a=p_a + i \epsilon_a \} ) } \nnel
&=&
\left(\frac{-1}{2}\right)^{N-1}
\left(\frac{1}{\pi}\right)^{N-l-1}
\dkj{N}  \; \delta( k_1+k_2+\ldots+k_N)
. \nnel &&
\permbnc \; \sum_{j=0}^{N-1} \sum_{ \{x\}\in X_j}
\twf (b_{x_0}b_{x_0+1}...b_{x_0-1};\{k\}) .
\nnel
&&
\frac{i \PP}{B^{N}_{x_0+1}} \ldots
\psi_{x_1}^{N+j} \delta(B^{N+j}_{x_1} ) \ldots
\psi_{x_l}^{N+j} \delta(B^{N+j}_{x_l} )
%\overline{ \frac{i \PP}{B^{N+j}_{2+j}} }
\ldots \frac{i \PP}{B^{N+j}_{N+j}}
\tselea{enptitfl2}
\end{eqnarray}
Here $\permbnc$ indicates that no cyclic permutations are included when
summing over the permutations of the $b$ indices, the
msum over different cycles of a given permutation being performed by the
$j$ sum.  $X_j$ was defined in \tref{eXdef}.

The next step is to try and sum over all cycles of the partitions, $\{ x
\}$, of the $b$
indices, and then to vary the partitions instead of
summing over all cycles of the partitions and then partitioning up the
indices into $l+1$ pieces.
Now define the label $k$ such that
$x_k - N$ is the smallest positive integer for a given set $\{ x \}$
i.e.\
\beq
 1 \leq x_k-N <
x_{k+1} < \ldots <x_l - N < x_0 < \ldots < x_{k-1} \leq N.
\eeq
We can relabel these $\{ x \}$
partitions using $\{ y_i \} $, ($i=0, \ldots, l$),
such that $y_0$ ($y_l$) is the smallest (biggest) in this series
i.e. $\{ y \} \in Y$ where
\beq
Y := \{ \{y \} | 1 \leq y_0 < y_1 < \ldots < y_l \leq N \}
\tseleq{eYdef}
\eeq
The definition is therefore
\bea
y_{i} = x_{i+k} - N &\mbox{if}& 0 \leq i \leq l-k
\nnel
y_{i} = x_{i+k-l-1} &\mbox{if}& l-k  < i \leq l      ,
\tselea{eydef}
\eea
so $y_{l-k+1} = x_0 =j+1$, $y_{l-k} = x_l \leq j+1+N$,
$y_0= x_k -N$ etc.  For each partition $\{ x \}$ and cycle $j$
we have a unique $\{ y \}$ and $k$ defined, \tref{eydef} is a 1:1
mapping from $\{ x \}$ to $ \{y \}$.

However we can now think of starting with a set of $l+1$
variables $\{ y \}$
which satisfy $1 \leq y_0 < \ldots < y_l \leq N$ and a variable $0\leq
k < N$.  The $k$ represents reverse cycling of the $l$ $y$'s that define a
partition of $1, \ldots, N$ (whereas $j$ represented cycling the $N$ $b$
indices).  We can then define a unique partition $\{ x \}$  and
variable $j$ through \tref{eydef} and this is a 1:1 map from
$\{ \{ y \} , k \} \rightarrow \{ \{ x \} , j \}$.
Thus this $\{ x \}$ to $ \{y \}$ map is
1:1 and onto and we can rewrite \tref{enptitfl2} as
\begin{eqnarray}
\lefteqn{\Phi_l(\{z_a=p_a + i \epsilon_a \} ) } \nnel
&=& \left(\frac{-1}{2}\right)^{N-1}
\left(\frac{1}{\pi}\right)^{N-l-1}
\dkj{N}  \; \delta( k_1+k_2+\ldots+k_N)
. \nnel &&
\permbnc \sum_{ \{y\} \in Y  } \sum_{k=0}^{l}
\twf (b_{x_0}b_{x_0+1}...b_{x_0-1}; \{k\}) .
\frac{i \PP}{B^{x_0-1}_{x_0+1}} \ldots
\psi_{x_1}^{x_0-1} \delta(B^{x_0-1}_{x_1} ) \ldots
\nnel
\\
&& \psi_{x_l}^{x_0-1} \delta(B^{x_0-1}_{x_l} )
\ldots \frac{i \PP}{B^{x_0-1}_{x_0-1}}
\tselea{enptitfl4}
\end{eqnarray}
where the $ \{ x \} $'s and $j$ are defined in terms of the $\set{y}$'s
and $k$ using \tref{eydef}.

Now the $\sum_k$ shifts the indices $l+1$ times, shifting by
$x_{k+1}-x_{k}$.  The reason for doing this rearrangement is in the
delta functions.
If we now set the external energies $\{ p \}$ to zero,
the integration variables associated with each
block sum to zero because the delta functions in \tref{enptitfl4} can
be written as
\beq
\delta( k_1+k_2+\ldots+k_N)
\prod_{j=1}^l \delta( B_{x_{j}}^{x_0-1} ) =
\prod_{j=1}^l\delta( \sum_{i=x_{j-1}}^{x_j} k_{b_i} )
\eeq
Together with the property \tref{etcyclee}, this shows that the same thermal
Wightman function factor appears, for a given partition \set{y},
however often the partition has been cycled.
In a similar way the delta functions ensure that the principal part
denominators can be written in a way that is independent of the $k$ sum.
This leaves us with
\begin{eqnarray}
\lefteqn{\Phi_l(\{z_a= 0 + i \epsilon_a \} ) } \nnel
&=& \left(\frac{-1}{2}\right)^{N-1}
\left(\frac{1}{\pi}\right)^{N-l-1}
\dkj{N}
\; \delta(k_1+k_2+\ldots+k_N) \;
. \nnel &&
\permbnc \sum_{ \{y\} \in Y  }
\twf (b_{y_0}b_{y_0+1}...b_{y_0-1}; \{k\}) .
\nnel &&
\frac{i \PP}{B(y_0+1,y_1-1; \{ k \} )}
\frac{i \PP}{B(y_0+2,y_1-1; \{ k \} )} \ldots \nnel
&&
\frac{i \PP}{B(y_1-1,y_1-1; \{ k \} )} .
\delta(B(y_1,y_2-1; \{ k \} ) ) \ldots
\nnel &&
\frac{i \PP}{B(y_1+1,y_2-1; \{ k \} )} \ldots
\frac{i \PP}{B(y_0-1,y_0-1; \{ k \} )} . \nnel
&& \Psi ({\psi})
\tselea{enptitfl5}
\eea
where the sum of cycles and
all the signs of the epsilon's are encoded in the
\bea
\Psi({\psi}) = \sum_{k=0}^{l}
\psi_{x_1}^{x_0-1} \ldots \psi_{x_l}^{x_0-1}
= \sum_{k=0}^l \psi_{y_{l-k+2}}^{y_{l-k+1}-1} \ldots
\psi_{y_{l-k}}^{y_{l-k+1}-1}  .
\tselea{ePsi}
\end{eqnarray}
This $\Psi$ is a generalisation of the expression \tref{e2ptPsi}
encountered in the case of two-point functions, and \tref{e3ptPsi1}
and \tref{e3ptPsi2} seen with three-point functions.
All dependence on the
precise analytic continuation, and hence the specification of
which GRF is being studied, is
contained in this factor.
We are now using the second definition of $B$ from \tref{eBdef2} for
convenience.

\subsection{The special case of the retarded and advanced functions}

For the case of retarded and advanced functions, the sum in $\Psi$ of
\tref{enptitfl5} is easy to perform.
Suppose we are looking at $R_a$ or $A_a$ so that $\epsilon_a$ is
of a different sign from all the others.  From \tref{eepcond2}
$\epsilon_a$ is then as large as the sum of the
other epsilons.  Then any sum of epsilons containing $\epsilon_a$ has
the same sign as $\epsilon_a$.  The remaining sums must have the
opposite sign.  Thus the product of $\psi$'s is equal to
\bea
\Psi=\sum_{k=0}^{l} (-\theta_a)^k (\theta_a)^{l-k}
%&=& (\theta_a)^l \sum_{k=0}^{l} (-1)^k
%\nnel
&=&
\protect\left\{
\begin{array}{cl}
0, &  l \mbox{ odd} \\
-1 , & l \mbox{ even}
\end{array}
\right.
\tselea{epsires1}
\eea
where $\theta_a=\pm 1$ depending on the sign of $\epsilon_a$.  Looking
at the terms with an even $l$, an even number of delta functions, we
see that they are independent of the $a$ label and equal
\begin{eqnarray}
\lefteqn{\Phi_l(\{z_a= 0 + i \epsilon_a \} ) } \nnel
&=& - \left(\frac{-1}{2}\right)^{N-1}
\left(\frac{1}{\pi}\right)^{N-l-1}
\dkj{N}
\; \delta(k_1+k_2+\ldots+k_N) \;
. \nnel &&
\permbnc \sum_{ \{y\} \in Y  }
\twf (b_{y_0}b_{y_0+1}...b_{y_0-1}; \{k\}) .
\nnel &&
\frac{i \PP}{B(y_0+1,y_1-1; \{ k \} )}
\frac{i \PP}{B(y_0+2,y_1-1; \{ k \} )} \ldots \nnel
&&
\frac{i \PP}{B(y_1-1,y_1-1; \{ k \} )} .
\delta(B(x_1,x_2-1; \{ k \} ) ) \ldots
\nnel &&
\frac{i \PP}{B(y_1+1,y_2-1; \{ k \} )} \ldots
\frac{i \PP}{B(y_0-1,y_0-1; \{ k \} )} .
\Psi ({\psi})
\tselea{enptitfl6}
\end{eqnarray}
and this is independent of the label $a$ chosen.
Hence
\bea
R_a(0) = A_a(0) &=&
\sum_{l=0,2,\ldots}^{N} \Phi_l = \mbox{
independent of }a
\eea
with the same number found for all values of the index $a$.  This shows that
all the zero energy retarded and advanced N-point thermal Green
functions made out of the same set of bosonic fields are equal at zero
energy.

\subsection{Generalised Retarded Functions}

The generalised retarded functions are the Green functions obtained
from any valid choice of epsilons, as specified by \tref{eepcond1} and
\tref{eepcond2}. Given that the retarded and advanced are
subsets of the generalised functions we therefore have to show that
\tref{epsires1} holds for all possible epsilon choices.  This can be
done as follows.

We first note that in \tref{ePsi} that the epsilons only appear in
blocks running from $y_j \ra (y_{j+1}-1)$.  It helps to simplify the
notation if one works in terms of such blocks.  So define
\bea
\epsilonbar_j &=& \sum_{m=y_j}^{(y_{j+1})-1} \epsilon_{b_m}
\\
\sigmabar_c^a =
\psi_{y_c}^{y_a-1} =  \sum_{l=0}^{a-c} \epsilonbar_{c+l}
&&
\psibar^a_c ={\rm SGN}(\sigmabar^a_c) = \psi_{y_c}^{y_{a+1}-1}
\\
\Lambda_k &=& \prod_{j=0}^{l-1} \psibar^k_{k-j}
\tselea{eLambdadef}
\\
\Psi&=&\sum_{k=0}^{l} \Lambda_k
\tselea{ePsidef2}
\eea
It is clear from the properties of the epsilons that the
$\set{\epsilonbar}$ also satisfy the same properties though with $N
\ra (l+1)$.  The idea of equivalent classes of $\epsilonbar$'s as
defined in \tref{eepequiv} is also of use, so we have
\bea
\set{\epsilonbar} \equiv \set{\dash{\epsilonbar}} & \mbox{ iff } &
{\rm SGN}( \sum_{j=1}^{j=N} c_j \epsilonbar_j) =
{\rm SGN}( \sum_{j=1}^{j=N} c_j \dash{\epsilonbar}{}_j ) \nnel
&& \forall \; \{c_j\}=\{0,1 \; | \; 0<\sum c_j <N\}
\tselea{eepbarequiv}
\eea
In this case though
each equivalence class of $\epsilonbar$'s does not pick  out
a unique $N$-point GRF (unless
$l+1=N$).

A useful lemma to prove is as follows:-

{\large\bf Lemma} {\em Each equivalence class of $\epsilonbar_j$
($j=0,\ldots,l$), as defined by \tref{eepbarequiv}, $\Lambda_k$ is a product of
a unique number of $+1$ and a unique number of $-1$.}

{\bf Proof}

Consider two different terms in  $\Psi$, $\Lambda_{k_1}$ and
$\Lambda_{k_2}$.
{}From the definition of the $\sigmabar$'s we have that
\beq
\sigmabar_{k_1-j_1}^{k_1} - \sigmabar_{k_1-j_2}^{k_1} =
\sigmabar_{k_1-j_1}^{k_1-j_2-1}
\eeq
where we will now take $j=0\ldots l$.  For the case of
$j_2=k_1-k_2-1$ we find a relation between terms appearing in
$\Lambda_{k_1}$ and
$\Lambda_{k_2}$ namely
\beq
\sigmabar_{k_1-j_1}^{k_1} - \sigmabar_{k_2+1}^{k_1} =
\sigmabar_{k_1-j_1}^{k_2}
\tseleq{epsibarrel}
\eeq
Now three different cases need to be considered, namely
\bea
\psibar_{k_1-j_1}^{k_1} \in \Lambda_{k_1} &\mbox{if}& j_1 \neq l-1
\tseleq{ecase1}
 \\
\psibar_{k_1-j_1}^{k_2}\in \Lambda_{k_2} &\mbox{if}& j_1 \neq k_1-k_2-1
\tseleq{ecase2}
\\
\psibar_{k_2+1}^{k_1} \in \Lambda_{k_1} &\mbox{as}& k_1 \neq k_2
\tseleq{ecase3}
\eea
where the indices are now periodic in $l+1$, $j_1 \equiv j_1+(l+1)$ etc.
The set $\Lambda$ is just the set of all the terms $\psibar$ appearing in
the product in the definition of $\Lambda$ \tref{eLambdadef}.

The indices $k_1$ and $k_2$ are fixed so that $\psibar_{k_2+1}^{k_1}
$ has a definite value.  Suppose $\psibar_{k_2+1}^{k_1} > 0 (<0) $.  Then
\begin{enumerate}
\item For the $(l-1)$
normal cases where $j_1 \neq l-1$ and $j_1 \neq k_1-k_2-1$ we see that
\end{enumerate}
\bea
\forall \; j_1 \neq l-1, j_1 \neq k_1-k_2-1 ,&&
\nnel
\psibar^{k_1}_{k_1-j_1} <0 (>0)
&\Rightarrow & \psibar_{k_1-j_1}^{k_2} <0 (>0) .
\tseleq{ecount}
\eea

Note that where $\psibar^{k_1}_{k_1-j_1} >0 (<0)$  nothing is learnt
from \tref{epsibarrel}.   Only one $\psibar$ term from each of
$\Lambda_{k_1}$ and  $\Lambda_{k_2}$ has  not considered in the above
$j_1$ range.  It follows from \tref{ecount} that apart from these last
terms,  the number of negative (positive) terms in the $\Lambda_{k_2}$
product is greater than or equal to the number of negative (positive)
terms in the  $\Lambda_{k_1}$ product.

The last term in each of the $\Lambda_{k_1}$ and  $\Lambda_{k_2}$
products, which are not covered by the above range of $j_1$,
correspond to the two cases $j_1 \neq l-1$ and $j_1 \neq k_1-k_2-1$.
In these cases one has trivial identities involving
$\psibar_{k_2+1}^{k_1} = - \psibar_{k_1+1}^{k_2} >0 (<0)$.  As these
last members of the $\Lambda$ products have these definite signs, we
therefore know that there must be more negative (positive) terms in
the $\Lambda_{k_2}$ product than in the  $\Lambda_{k_1}$ product.

Each term in the $\Lambda$ product is either positive or negative
(never zero from \tref{eepcond1}) and there are the same number of
terms in all the products.  So there must be fewer positive
(negative) terms in the $\Lambda_{k_2}$ product than in the
$\Lambda_{k_1}$ product.  Thus it is clear that the number of
positive and negative terms in the two $\Lambda_{k_1}$ and
$\Lambda_{k_2}$ products must both be different whatever the sign of
$\psibar^{k_1}_{k_1-j_1} $ is.

Since $k_1$ and $k_2$ are arbitrary, each $\Lambda_k$ term must
therefore have a unique number of positive and a unique number of
negative terms in its product. {\bf Q.E.D.}

{\large\bf Theorem} {\em For any allowed set of $\epsilonbar_j$
($j=0,\ldots,l$) as defined by \tref{eepcond1} and \tref{eepcond2},
$\Psi$ of
\tref{ePsidef2} is $0$ if $l$ is odd, $-1$ if $l$ is even.}

{\bf Proof}

{}From \tref{ePsidef2}, $\Psi$ is the sum of $(l+1)$ different
$\Lambda_k$'s.  Each $\Lambda_k$ is a product of $l$ terms, so  there
are $(l+1)$ different combinations of positive and negative terms each
$\Lambda_k$ can be,  from $0$ positive and $(l+1)$ negative through to
$(l+1)$ positive and $0$ negative terms (none of them zero by
\tref{eepcond2}).   From the lemma proved above, each of the  $(l+1)$
$\Lambda_k$'s is a product of a unique number of positive and negative
terms.  Hence each of the $(l+1)$ different combinations of $l$ positive
and negative terms in the $\Lambda$'s, $l$ term products, appear
once and only once in the $\Psi$ sum.  Thus
\beq
\Psi = \sum_{k=0}^{l} (+1)^{l+1-k}(-1)^k
=\protect\left\{
\begin{array}{cl}
0, & l \mbox{ odd} \\
-1 , &  l \mbox{ even}
\end{array}
\right.
{\rm\bf Q.E.D.}
\eeq

Thus we have shown that the GRF at zero energy are all equal.  The
N-point retarded and advanced function results hold for all N-point GRF
namely
\begin{eqnarray}
\Phi ( \{ 0+i\epsilon_a \} ) &=&
\sum_{l=0,2,\ldots}^{N} \Phi_l \; \; \mbox{
independent of} \{ \epsilon \} \mbox{ chosen}
\eea
\bea
\lefteqn{\Phi_l(\{z_a= 0 + i \epsilon_a \} ) } \nnel
&=& - \left(\frac{-1}{2}\right)^{N-1}
\left(\frac{1}{\pi}\right)^{N-l-1}
\dkj{N}
\; \delta(k_1+k_2+\ldots+k_N) \;
. \nnel &&
\permbnc \sum_{ \{y\} \in Y  }
\twf (b_{y_0}b_{y_0+1}...b_{y_0-1};\{k\}) .
\nnel &&
\frac{i \PP}{B(y_0+1,y_1-1; \{ k \} )}
\frac{i \PP}{B(y_0+2,y_1-1; \{ k \} )} \ldots \nnel
&&
\frac{i \PP }{B(y_1-1,y_1-1; \{ k \} )} .
\delta(B(y_1,y_2-1; \{ k \} ) ) \ldots
\nnel &&
\frac{i \PP }{B(y_1+1,y_2-1; \{ k \} )} \ldots
\frac{i \PP }{B(y_0-1,y_0-1; \{ k \} )} .
\nnel
&=& \mbox{ independent of } \{ \epsilon \} \mbox{ chosen}
\tselea{enptitfl7}
\end{eqnarray}

Hence, at the point where all external energies are zero, {\em all}
bosonic generalised retarded functions are equal.   The discontinuities
across the cuts in these functions are zero at this point.

\section{Conclusions}

We have shown that the discontinuities across the cuts in the $\Phi$
function of \tref{ePhidef}, which contains all the various possible
GRF made out of a single set of bosonic fields, are zero.  This
answers several of the questions raised in the introduction.

For derivative expansions of effective actions,  it shows that the lowest
order term (which includes the free energy or effective potential) is
the same which ever side of the energy cuts one starts from.
However one should expect differences to appear in higher order terms
of a derivative expansion as we have only shown that the
discontinuities are zero precisely at the zero energy point, and not
in a neighbourhood of the zero energy point.  Conversely it will not
matter which GRF one uses to calculate the lowest order term of the
derivative expansions.  This is one reason why there is never any
discussion of analytic continuation in an ITF calculation of free
energies.  The lack of discontinuities at zero energy is also the
technical reason why the free energy can be real.

The result also means that the discrete zero energy point of the set
of Euclidean ITF energy values, $\{ z_j = 2 \pi i \nu_j / \beta\}$,
can be thought of as lying on any side of the of any of the cuts
without any inconsistency.  Hence, a Green function evaluated at
this discrete point can be thought of as being related by analytic
continuation to any one of the GRF.  It also means that one can turn
a sum over discrete energies into integrals along the real energy
axis in the usual way \tcite{LvW} without any ambiguity arising from
the position of the zero energy point relative to the cuts.  This
technicality is not usually noted in the discussion of the use of
such methods for doing ITF energy sums.

\section*{Acknowledgements}

I would like to thank M.\ van Eijck, F.\ Gu\'{erin} and R.\ Kobes for
discussions, the University of McGill for hospitality
while some of this work was done,
and the Royal Society for their support through a University
Research Fellowship.

\npagepub

%Bibliography for ze.tex
\typeout{ZE - references}

%Example of citation:-
%  Schwinger and Keldysh \cite{sch 1961,kel 1965}

\npagepub
%\typeout{--- Figure at end for Submission ---}

\end{document}